\documentclass[11pt]{article}
\usepackage{graphicx}
\usepackage{authblk}
\usepackage{hyperref}

\usepackage[numbers]{natbib}
\usepackage{amsmath}
\usepackage{url}

\begin{document}

\title{Representations of Materials for Machine Learning\footnote{Accepted for publication in Annual Review of Materials Research Volume 53, https://www.annualreviews.org/.}}


\author[1]{James Damewood}
\author[1,2]{Jessica Karaguesian}
\author[1]{Jaclyn R. Lunger}
\author[1]{Aik Rui Tan}
\author[1,3]{Mingrou Xie}
\author[1]{Jiayu Peng}
\author[1]{Rafael G\'omez-Bombarelli\thanks{rafagb@mit.edu}}

\affil[1]{Department of Materials Science and Engineering, Massachusetts Institute of Technology, 77 Massachusetts Avenue, Cambridge, USA, 02129}
\affil[2]{Center for Computational Science and Engineering, Massachusetts Institute of Technology, 77 Massachusetts Avenue, Cambridge, MA, USA, 02139}
\affil[3]{Department of Chemical Engineering, Massachusetts Institute of Technology, 77 Massachusetts Avenue, Cambridge, MA, USA, 02139}


\maketitle

\begin{abstract}
High-throughput data generation methods and machine learning (ML) algorithms have given rise to a new era of computational materials science by learning relationships among composition, structure, and properties and by exploiting such relations for design. However, to build these connections, materials data must be translated into a numerical form, called a representation, that can be processed by a machine learning model. Datasets in materials science vary in format (ranging from images to spectra), size, and fidelity. Predictive models vary in scope and property of interests. Here, we review context-dependent strategies for constructing representations that enable the use of materials as inputs or outputs of machine learning models. Furthermore, we discuss how modern ML techniques can learn representations from data and transfer chemical and physical information between tasks. Finally, we outline high-impact questions that have not been fully resolved and thus, require further investigation. 
\end{abstract}


\tableofcontents

\section{INTRODUCTION}\label{intro}

Energy and sustainability applications demand the rapid development of scalable new materials technologies. Big data and machine learning (ML) have been proposed as strategies to rapidly identify ``needle-in-the-haystack" materials that have the potential for revolutionary impact.

High-throughput experimentation platforms based on robotized laboratories can increase the efficiency and speed of synthesis and characterization. However, in many practical open problems, the number of possible design parameters is too large to be analyzed exhaustively. Virtual screening somewhat mitigates this challenge by using physics-based simulations to suggest the most promising candidates, reducing the cost but also the fidelity of the screens\cite{GomezBombarelli2018,Peng2022,Pyzer-Knapp2015}.

Over the past decade, hardware improvements, new algorithms, and the development of large-scale repositories of materials data \cite{Jain2013,Kirklin2015,Chanussot2021,Curtarolo2012,Ward2018,Tran2022} have enabled a new era of ML methods. In principle, predictive ML models can identify and exploit nontrivial trends in high-dimensional data to achieve accuracy comparable with or superior to first-principles calculations, but with orders of magnitude reduction in cost. In practice, while a judicious model choice is helpful in moving towards this ideal, such ML methods are also highly dependent on the numerical inputs used to describe systems of interest---the so-called representations. Only when the representation is composed of a set of features and descriptors from which the desired physics and chemistry are emergent can the promise of ML be achieved.

Thus, the problem that materials informatics researchers must answer is: how can we best construct this representation? Previous works have provided practical advice for constructing materials representations \cite{Huo2017,Faber2015,Bartok2013,Lilienfeld2015,Musil2021}, namely that: (1) the similarity/difference between two data points should match the similarity/difference between representations of those two data points, (2) the representation should be applicable to the entire materials domain of interest, (3) the representation should be easier to calculate than the target property. 

Representations should reflect the degree of similarity between data points such that similar data have similar representations and as data points become more different their representations  diverge. Indeed, the definition of similarity will depend on the application. Consider, as an example, a hypothetical model predicting the electronegativity of an element, excluding Nobel gases. One could attempt to train the model using atomic number as input, but this representation violates the above principle, as atoms with a similar atomic number can have significantly different electronegativities (e.g. fluorine and sodium), forcing the model to learn a sharply varying function whose changes appear at irregular intervals. Alternatively, a representation using period and group numbers would closely group elements with similar atomic radii and electron configurations. Over this new domain, the optimal prediction will result in a smoother function that is easier to learn. 

The approach used to extract representation features from raw inputs should be feasible over the entire domain of interest---all data points used in training and deployment. If data required to construct the representation is not available for a particular material, ML screening predictions cannot be made. 

Finally, for the ML approach to remain a worthwhile investment, the computational cost of obtaining representation features and descriptors for new data should be smaller than that of obtaining the property itself through traditional means, either experimentally or with first-principles calculations. If, for instance, accurately predicting a property calculated by density functional theory (DFT) with ML requires input descriptors obtained from DFT on the same structure and at the same level of theory, the machine learning model does not offer any benefit.

A practicing materials scientist will notice a number of key barriers to forming property-informative representations that satisfy these criteria. First, describing behavior often involves quantifying structure-to-property relationships across length scales. The diversity of possible atomistic structure types considered can vary over space groups, supercell size, and disorder parameters. This challenge motivates researchers to develop flexible representations capable of capturing local and global information based on atomic positions. Beyond this idealized picture, predicting material performance relies upon understanding the presence of defects, the characteristics of the microstructure, and reactions at interfaces. Addressing these concerns requires extending previous notions of structural similarity or developing new specialized tools. Furthermore, atomistic structural information is not available without experimental validation or extensive computational effort \cite{Glass2006,Wang2010}. Therefore, when predictions are required for previously unexplored materials, models must rely on more readily available descriptors such as those based on elemental composition and stoichiometry. Lastly, due to experimental constraints, datasets in materials science can often be scarce, sparse, and restricted to relatively few and self-similar examples. The difficulty in constructing a robust representation in these scenarios has inspired strategies to leverage information from high-quality representations built for closely related tasks through transfer learning.

\begin{figure}[h]
\includegraphics[width=6in]{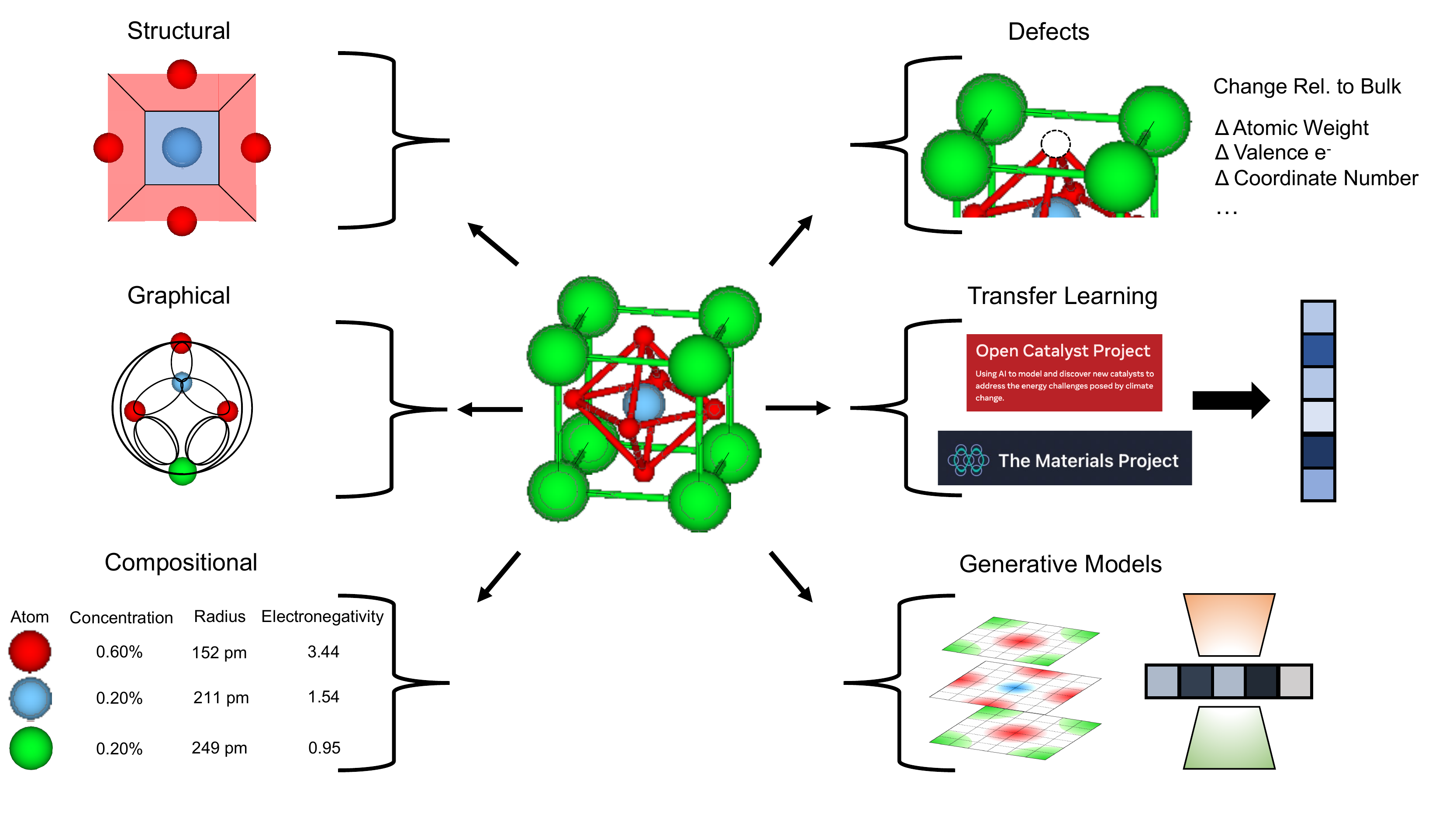}
\caption{Summary of representations for perovskite SrTiO$_{3}$. \textbf{Top Left.} 2D cross section of Voronoi decomposition. Predictive features can be constructed from neighbors and geometric shape of cells \cite{Ward2017}. \textbf{Middle Left.} Crystal graph of SrTiO$_{3}$ constructed assuming periodic boundary conditions and used as input to graph neural networks \cite{Xie2018}. \textbf{Bottom Left.} Compositional data including concentrations and easily accessible atomic features including electronegativities and atomic radii \cite{Ward2016}. Data taken from Reference \cite{Handbook}. \textbf{Top Right.} Deviations on a pristine bulk structure induced by an oxygen vacancy to predict formation energy \cite{Frey2020}. \textbf{Middle Right.} Representations can be learned from large repositories using deep neural networks. The latent physical and chemical information can be leveraged in related but data-scare tasks. \textbf{Bottom Right.} Training of generative models capable of proposing new crystal structures by placing atoms in discretized volume elements \cite{Hoffmann2019,Noh2019,Court2020,Choubisa2020}.}
\label{summary}
\end{figure}

In this review, we will analyze how representations of solid-state materials (\textbf{Figure \ref{summary}}) can be developed given constraints on the format, quantity, and quality of available data. We will discuss the justification, benefits, and trade-offs of different approaches. This discussion is meant to highlight methods of particular interest rather than provide exhaustive coverage of the literature. We will discuss current limitations and open problems whose solutions would have high impact. In summary, we intend to provide readers with an introduction to current state of the field and exciting directions for future research.

\section{STRUCTURAL FEATURES FOR ATOMISTIC GEOMETRIES}

Simple observations in material systems (e.g. higher ductility of face-centered cubic metals compared to body-centered cubic metals) have made it evident that material properties are highly dependent on crystal structure---from coordination and atomic ordering to broken symmetries and porosity. For a computational material scientist, this presents the question of how to algorithmically encode information from a set of atoms types (${a_1,a_2, a_3, ...}$), positions (${x_1,x_2, x_3, ...}$), and primitive cell parameters into a feature set that can be effectively utilized in machine learning. 

For machine learning methods to be effective, it is necessary that the machine-readable representation of a material's structure fulfills the criteria as outlined in the introduction \cite{Huo2017,Faber2015,Bartok2013,Lilienfeld2015,Musil2021}. Notably, scalar properties (such as heat capacity or reactivity) do not change when translations, rotations, or permutations of atom indexing are applied to the atomic coordinates. Therefore, to ensure  representations reflect the similarities between atomic structures,  the representations should also be invariant to those symmetry operations.

\subsection{Local Descriptors} \label{subsec:local}

\begin{figure}[h]
\includegraphics[width=6in]{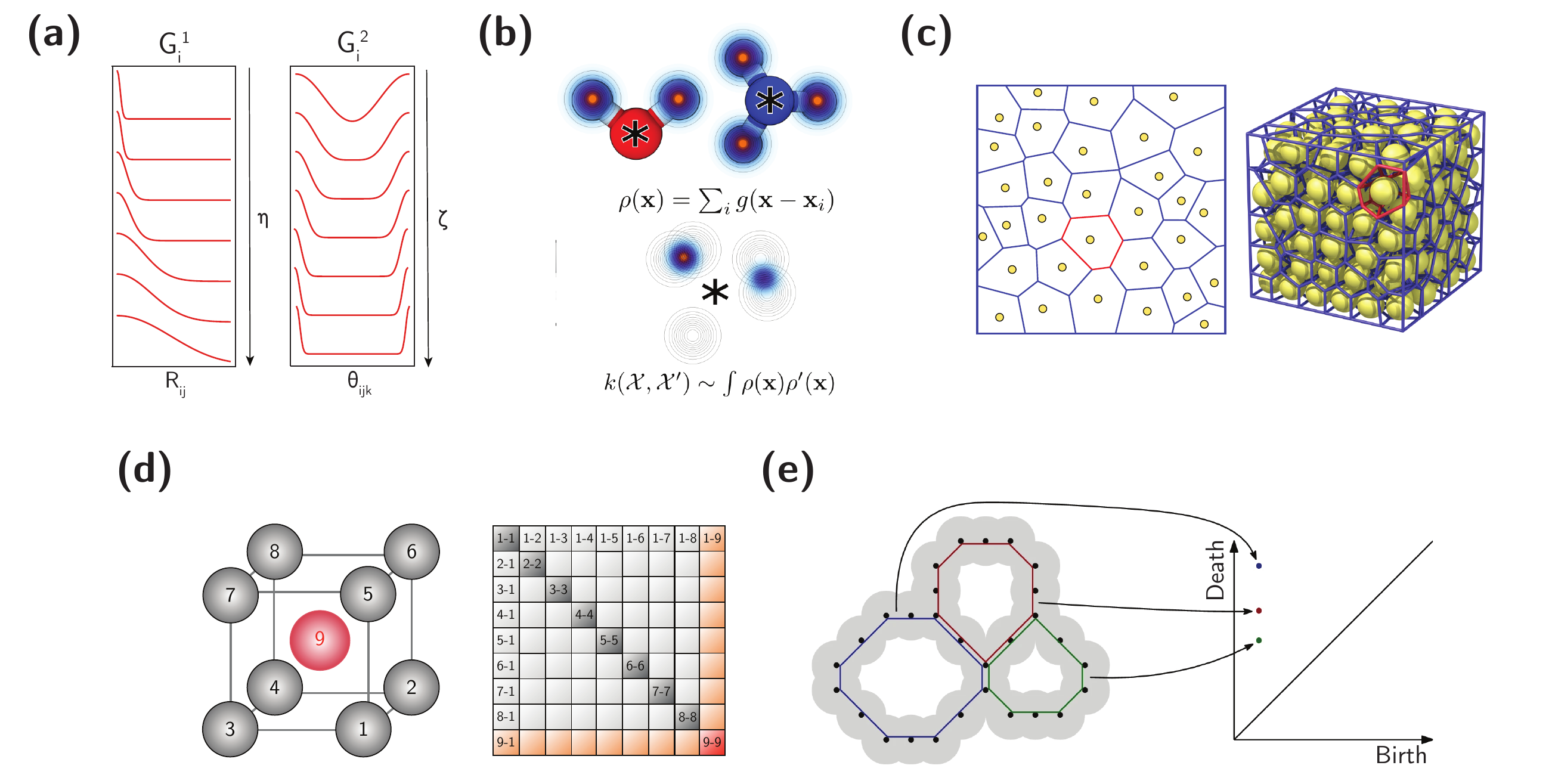}
\caption{
\textbf{(a)} Examples of radial, $G^1_i$, and angular, $G^2_i$, symmetry functions from the local atom-centered symmetry function descriptor proposed by Behler and Parinello\cite{Behler2007}.
\textbf{(b)} In the Smooth Overlap of Atomic Positions (SOAP) descriptor construction, the atomic neighborhood density of a central atom is defined by a sum of Gaussian functions around each neighboring atom. A kernel function can then be built to compare the different environments by computing the density overlap of the atomic neighborhood functions. Figure is reprinted from reference \cite{Bartok2017}.
\textbf{(c)} Voronoi tessellation in two and three dimensions. Yellow circles and spheres show particles while the blue lines divide equidistantly the space between two neighboring particles. Polygonal spaces encompassed by the blue lines are the Voronoi cells. Figure is reprinted from reference \cite{Lazar2022}.
\textbf{(d)} Illustration of a Coulomb matrix where each element in the matrix shows Coulombic interaction between the labeled particles in the system on the left. Diagonal elements show self-interactions. 
\textbf{(e)} The births and deaths of topological holes in a point cloud (left) are recorded on a persistence diagram (right). Persistent features lie far from the parity line and indicate more significant topological features. Feature is reprinted from reference \cite{Krishnapriyan2021}.
}
\label{structural}
\end{figure}

One strategy to form a representation of a crystal structure is to characterize the local environment of each atom and consider the full structure as a combination local representations. This concept was applied by Behler and Parinello\cite{Behler2007}, who proposed the atom-centered symmetry functions (ACSF). ACSF descriptors (\textbf{Figure \ref{structural}a}) can be constructed using radial, $G_i^1$, and angular, $G_i^2$, symmetry functions centered on atom \textit{i},
\begin{equation}
    G_{i}^1 = \sum^{\text{neighbors}}_{j \ne i} e^{-\eta (R_{ij} - R_s)^2} f_c(R_{ij})
\end{equation}
\begin{equation}
    G_{i}^2 = 2^{1-\zeta} \sum^{\text{neighbors}}_{j, k \ne i} (1 + \lambda \cos \theta_{ijk})^\zeta  e^{-\eta (R_{ij}^2 + R_{ik}^2 + R_{jk}^2)} f_c(R_{ij}) f_c(R_{ik}) f_c(R_{jk})
\end{equation}
with the tunable parameters $\lambda$, $R_{s}$, $\eta$, and $\zeta$. $R_{ij}$ is the distance between the central atom $i$ and atom $j$, and $\theta_{ijk}$ corresponds to the angle between the vector from the central atom to atom $j$ and the vector from the central atom to atom $k$. The cutoff function $f_c$ screens out atomic interactions beyond a specified cutoff radius and ensures locality of the atomic interactions. Because symmetry functions rely on relative distances and angles, they are rotationally and translationally invariant. Local representations can be constructed from many symmetry functions of the type $G_{i}^1$ and $G_{i}^2$ with multiple settings of tunable parameters to probe the environment at varying distances and angular regions. With the set of localized symmetry functions, neural networks can then predict local contributions to a particular property and approximate global properties as the sum of local contributions. The flexibility of this approach allows for modification of the $G_i^1$ and $G_i^2$ functions \cite{Behler2011,Smith2017} or higher capacity neural networks for element-wise prediction \cite{Smith2017}. 

 In search of a representation with fewer hand-tuned parameters and a more rigorous definition of similarity, Bartok et al. \cite{Bartok2013} proposed a rotationally invariant kernel for comparing environments based on local atomic density. Given a central atom, the Smooth Overlap of Atomic Positions (SOAP) defines the atomic density function $\rho(\mathbf{r})$ as a sum of Gaussian functions centered at each neighboring atom within a cutoff radius (\textbf{Figure \ref{structural}b}). The choice of Gaussian function is motivated by the intuition that representations should be continuous such that small changes in atomic positions should result in correspondingly small changes in the metric between two configurations. With a basis of radial functions $g_n(\mathbf{r})$ and spherical harmonics $Y_{lm}(\theta, \phi)$, $\rho(\mathbf{r})$ for central atom $i$ can be expressed as:
\begin{equation}
      \rho_{i}(\mathbf{r}) = \sum_j \exp{-\frac{|\mathbf{r} - \mathbf{r}_{ij}|^2}{2\sigma^2}} = \sum_{nlm} c_{nlm} g_n(\mathbf{r}) Y_{lm}(\mathbf{\hat{r}})
\end{equation}

and the kernel can be computed\cite{Bartok2013,De2016}:

\begin{equation}
    K(\rho, \rho') = \mathbf{p}(\mathbf{r}) \cdot \mathbf{p}'(\mathbf{r})
\end{equation}
\begin{equation}
    \mathbf{p}(\mathbf{r}) \equiv \sum_m c_{nlm}(c_{n'lm})^{*}
\end{equation}

where $c_{nlm}$ are the expansion coefficients in \textbf{Equation 3}. In practice, $\mathbf{p}(\mathbf{r})$ can be used as a vector descriptor of the local environment and is also referred to as a power spectrum \cite{Bartok2013}.
SOAP has demonstrated extraordinary versatility for materials applications both as a tool for measuring similarity \cite{Schwalbe-Koda2019} and as a descriptor for machine learning algorithms \cite{Dragoni2018}. Furthermore, the SOAP kernel can be used to compare densities of different elements by adding an additional factor that provides a definition for similarity between atoms, where for instance, atoms in the same group could have higher similarity \cite{De2016}. The mathematical connections between different local atomic density representations including ACSFs and SOAP are elucidated by a generalized formalism introduced by Willatt et al. \cite{Willatt2019}, offering a methodology through which the definition of new variants can be clarified.

Instead of relying on the density of nearby atoms, local representations can be derived from a Voronoi tessellation of a crystal structure. The Voronoi tessellation segments space into cells such that each cell contains one atom and all points in space such that atom A is the closest atom are contained in the same cell as atom A (\textbf{Figure \ref{structural}c}). From these cells, Ward et al. \cite{Ward2017} identified a set of descriptive features including an effective coordination number computed using the area of the faces, the lengths and volumes of nearby cells, ordering of the cells based on elements, and atomic properties of nearest neighbors weighted by the area of the intersecting face. When combined with compositional features \cite{Ward2016}, their representation results in better performance on predictions of formation enthalpy for ICSD than partial radial distribution functions \cite{Schutt2014Arx} (\textbf{Figure 1} in Reference \cite{Ward2017}). In subsequent work, these descriptors have facilitated the prediction of experimental heat capacities in MOFs \cite{Moosavi2022}. Similarly, Isayev et al. \cite{Isayev2017} replaced faces of the Voronoi tessellation with virtual bonds and separated the resulting framework into sets of linear (up to four atoms) and shell-based (up to nearest neighbors) fragments. Additional features related to the atomic properties of constituent elements were associated with each fragment, and the resulting vectors were concatenated with attributes of the supercell. In addition to demonstrating accurate predictive capabilities, models could be interpreted through the properties of the various fragments. For instance, predictions of band gap could be correlated with the difference in ionization potential in two-atom linear fragments, a trend that could be exploited to design material's properties through tuning of composition\cite{Isayev2017}.

\subsection{Global Descriptors}

Alternatively, to more explicitly account for interactions beyond a fixed cutoff, atom types and positions can be encoded into a global representation that reflects geometric and physical insight. Inspired by the importance of electrostatic interactions in chemical stability, Rupp et al. \cite{Rupp2012} proposed the Coulomb matrix (\textbf{Figure \ref{structural}d}), which models the potential between electron clouds:

\begin{equation}
    M_{i,j} = 
    \begin{cases}
        Z_i^{2.4} & \text{for } i=i\\
        \frac{Z_i Z_j}{|r_i - r_j|} & \text{for } i \neq j
    \end{cases}
\end{equation}

Due to the fact that off-diagonal elements are only dependent on relative distances, Coulomb matrices are rotation and translation invariant. However, the representation is not permutation invariant since changing the labels of the atoms will rearrange the elements of the matrix. While originally developed for molecules, the periodicity of crystal structures can be added to the representation by considering images of atoms in adjacent cells, replacing the $\frac{1}{|r_i - r_j|}$ dependence with another function with the same small distance limit and periodicity that matches the parent lattice, or using an Ewald sum to account for long range interactions \cite{Faber2015}. BIGDML \cite{Sauceda2022} further improved results by restricting predictions from the representation to be invariant to all symmetry operations within the space group of the parent lattice and demonstrated effective implementations on tasks ranging from H interstitial diffusion in Pt to phonon density of states. While this approach has been able to effectively model long-range physics, these representations rely on a fixed supercell and may not be able to achieve the same chemical generality as local environments \cite{Sauceda2022}. 

Global representations have also been implemented with higher-order tensors. Partial radial distribution functions (PRDF) are 3D non-permutation invariant matrices $g_{\alpha\beta r}$ whose elements correspond to the density of element $\beta$ in the environments of element $\alpha$ at radius $r$ \cite{Schutt2014Arx}. The many-body tensor representation (MBTR) provides a more general framework \cite{Huo2017} that can quantify k-body interactions and account for chemical similarity between elements. The MBTR is translationally, rotationally, and permutation invariant and can be applied to crystal structures by only summing over atoms in the primitive cell. While MBTR exhibited better performance than SOAP or Coulomb matrices for small molecules, its accuracy may not extend to larger systems \cite{Huo2017}.

Another well-established method for representing crystal structures in materials science is the cluster expansion. Given a parent lattice and a decoration $\mathbf{\sigma}$ defining the element that occupies each site, Sanchez et al. sought to map this atomic ordering to material properties and proposed evaluating the correlations between sites through a set of cluster functions. Each cluster function $\Phi$ is constructed from a product of basis functions ${\phi}$, over a subset of sites \cite{Sanchez1984}. To ensure the representation is appropriately invariant, symmetrically equivalent clusters are grouped into classes denoted by $\alpha$. The characteristics of the atomic ordering can be quantified by averaging cluster functions over the decoration $<\Phi_{\alpha}>_{\mathbf{\sigma}}$, and properties $q$ of the configuration can be predicted as:

\begin{equation}
    q(\mathbf{\sigma}) = \sum_{\alpha}J_{\alpha}m_{\alpha}<\Phi_{\alpha}>_{\sigma}
\end{equation}

where $m_{\alpha}$ is a multiplicity factor that accounts for the rate of appearance of different cluster types, and $J_{\alpha}$ are parameters referred to as effective cluster interactions that must be determined from fits to data \cite{Chang2019}. While cluster expansions have been constructed for decades and provided useful models for configurational disorder and alloy thermodynamics \cite{Chang2019}, cluster expansions assume the structure of the parent lattice and models cannot generally be applied across different crystal structures \cite{Hart2021,Nyshadham2019}. Furthermore, due to the increasing complexity of selecting cluster functions, implementations are restricted to binary and ternary systems without special development \cite{Yang2022}. Additional research has extended the formalism to continuous environments (Atomic Cluster Expansion) by treating $\mathbf{\sigma}$ as pairwise distances instead of site-occupancies and constructing ${\phi}$ from radial functions and spherical harmonics \cite{Drautz2019}. The Atomic Cluster Expansion framework has provided a basis for more sophisticated deep learning approaches \cite{Batatia2022}.

\subsection{Topological Descriptors}

Topological data analysis (TDA) has found favor over the past decade in characterizing structure in complex, high-dimensional datasets. When applied to the positions of atoms in amorphous or crystalline structures, topological methods reveal underlying geometric features that inform behavior in downstream predictions such as phase changes, reactivity, and separations. In particular, persistent homology (PH) is able to identify significant structural descriptors that are both machine readable and physically interpretable. The data can be probed at different length scales (formally called filtrations) by computing a series of complexes that each include all sets of points where all pairwise distances are less than the corresponding length \cite{Carlsson2020}. Analysis of complexes by homology in different dimensions reveals holes or voids in the data manifold, which can be described by the range of length scales they are observed at (persistences), as well as when they are produced (births) and disappear (deaths). Emergent features with significant persistence values are less likely to be caused by noise in the data or as an artifact of the chosen length scales. In practice, multiple persistences, births, and deaths produced from a single material can be represented together by persistent diagrams (\textbf{Figure \ref{structural}e}) or undergo additional feature engineering to generate a variety of descriptors as machine learning inputs \cite{Pun2022}.

While persistent homology has been applied to crystal structures in the Open Quantum Material Database \cite{Jiang2021}, the method is particularly useful in the analysis of porous materials. The identified features (births, deaths, persistences) hold direct physical relevance to traditional structural features used to describe the pore geometries. For instance, persistent 2D deaths represent the largest sphere that can be included inside the pores of the materials. Krishnapriyan et al. has showed that these topological descriptors outperform traditional structural descriptors when predicting carbon dioxide adsorption under varying conditions for metal-organic frameworks \cite{Krishnapriyan2021}, as did Lee et al. for zeolites for methane storage capacities \cite{Lee2018}. Representative cycles can trace the topological features back to the atoms that are responsible for the hole or the void, creating a direct relationship between structure and predicted performance (\textbf{Figure 4} in Reference \cite{Krishnapriyan2021}). Similarity methods for comparing barcodes can then be used to identify promising novel materials with similar pore geometries for targeted applications. 

A caveat is that PH does not inherently account for system size and is thus size-dependent. The radius cutoff, or the supercell size, needs to be carefully considered to encompass all significant topological features and allow comparison across systems of interest. In the worst case scenario, the computation cost per filtration for a structure is $O(N^3)$, where $N$ is the number of sets of points defining a complex. Although the cost is alleviated by the sparsity of the boundary matrix \cite{Buchet2018}, the scaling is poor for structures whose geometric features exceed unit cell lengths. The benefit of using PH features to capture more complex structural information has to be carefully balanced with the cost of generating these features.

\section{LEARNING ON PERIODIC CRYSTAL GRAPHS}

In the previous section, we described many physically-inspired descriptors that characterize materials and can be used to efficiently predict properties. The use of differentiable graph-based representations in convolutional neural networks, however, mitigates the need for manual engineering of descriptors \cite{Duvenaud2015,Reiser2022}. Indeed, advances in deep learning and the construction of large-scale materials databases \cite{Jain2013,Kirklin2015,Chanussot2021,Curtarolo2012,Ward2018,Tran2022} have made it possible to learn representations directly from structural data. From a set of atoms ${a_1,a_2,...}$ located at positions ${x_1,x_2, x_3, ...}$, materials can be converted to a graph $G(V,E)$ defined as the set of atomic nodes $V$ and the set of edges $E$ connecting neighboring atoms. Many graph-based neural network architectures were originally developed for molecular systems, with edges representing bonds. By considering periodic boundary conditions and defining edges as connections between neighbors within a cutoff radius, graphical representations can be leveraged for crystalline systems. The connectivity of the crystal graph thus naturally encodes local atomic environments \cite{Xie2018}.

\begin{figure}[h]
\includegraphics[width=6in]{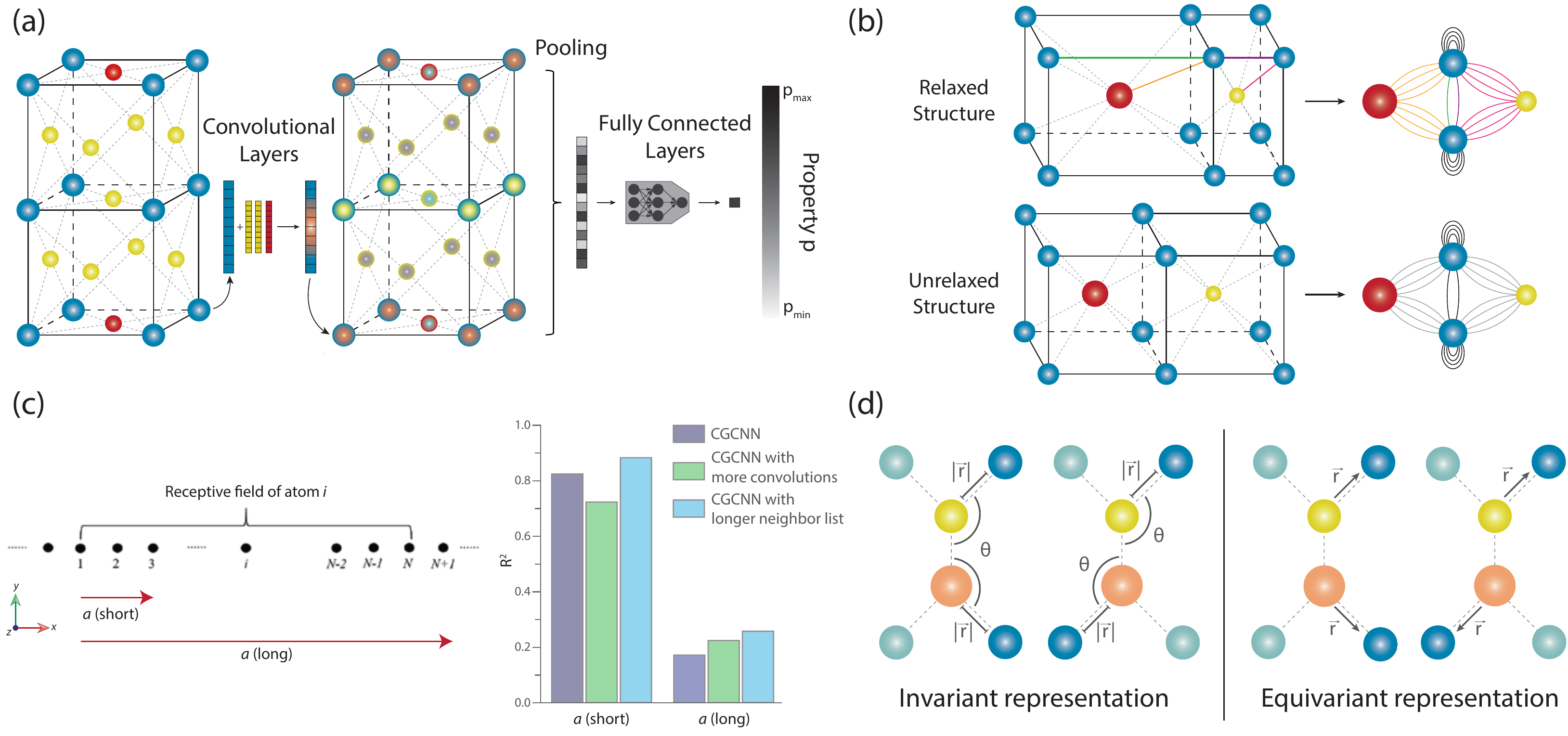}
\caption{
\textbf{(a)} General architecture of graph convolutional neural networks for property prediction in crystalline systems. Three-dimensional crystal structure is represented as a graph with nodes representing atoms and edges representing connections between nearby atoms. Features (e.g. nodal, edge, angular) within local neighborhoods are convolved, pooled into a crystal-wide vector, then mapped to the target property. Figure adapted from \cite{Karaguesian2021}.
\textbf{(b)} Information loss in graphs built from pristine structures. Geometric distortions of ground-state crystal structures are captured as differing edge features in graphical representations. This information is lost in graphs constructed from corresponding unrelaxed structures.
\textbf{(c)} Graph-based models can struggle to capture periodicity-dependent properties, such as cell lattice parameters. $R^2$ scores presented here were reported by Gong et al. for lattice parameter, \textit{a}, predictions in short and long 1D single carbon chain toy structures. Figure adapted from \cite{Gong2022}.
\textbf{(d)} Ability of graphical representations to distinguish toy structures. Assuming a sufficiently small cutoff radius, the invariant representation---using edge lengths and/or angles---cannot distinguish the two toy arrangements, while  the equivariant representation with directional features can. Figure adapted from \cite{Painn2021}.
}
\label{graphical}
\end{figure}

When used as input to machine learning algorithms, the graph nodes and edges are initialized with an associated set of features. Nodal features can be as simple as a one-hot vector of the atomic number or can explicitly include other properties of the atomic species (e.g. electronegativity, group, period). Edge features are typically constructed from the distance between the corresponding atoms. Subsequently, a series of convolutions parameterized by neural networks modify node and/or edge features based on the current state of their neighborhood (\textbf{Figure \ref{graphical}a}). As the number of convolutions increases, interactions from further away in the structure can propagate, and graph features become tuned to reflect the local chemical environment. Finally, node and edge features can be pooled to form a single vector representation for the material \cite{Duvenaud2015,Schnet2018}.

Crystal Graph Convolution Neural Networks (CGCNN) \cite{Xie2018} and Materials Graph Networks (MEGNet) \cite{Chen2019} have become benchmark algorithms capable of predicting properties across solid-state materials domains including bulk, surfaces, disordered systems, and 2D materials \cite{Fung2021a,Chen2021b}. Atomistic Line Graph Neural Network (ALIGNN) extended these approaches by including triplet three-body features in addition to nodes and edges and exhibited superior performance to CGCNN over a broad range of regression tasks including formation energy, bandgap, and shear modulus \cite{Choudhary2021}. Other variants have used information from Voronoi polyhedra to construct graphical neighborhoods and augment edge features \cite{Park2020} or initialized node features based on the geometry and electron configuration of nearest neighbor atoms \cite{Karamad2020}. 

While these methods have become widespread for property prediction, graph convolution updates based only on the local neighborhood may limit the sharing of information related to long-range interactions or extensive properties. Gong et al. demonstrated that these models can struggle to learn materials properties reliant on periodicity, including characteristics as simple as primitive cell lattice parameters (\textbf{Figure \ref{graphical}c})\cite{Gong2022}. As a result, while graph-based learning is a high-capacity approach, performance can vary substantially by the target use case. In some scenarios, methods developed primarily for molecules can be effectively implemented ``out-of-the-box" with the addition of periodic boundary conditions, but especially in the case of long-range physical phenomena, optimal results can require specialized modeling.

Various strategies to account for this limitation have been proposed. Gong et al. found that if the pooled representation after convolutions was concatenated with human-tuned descriptors, errors could be reduced by $90\%$ for related predictions, including phonon internal energy and heat capacity \cite{Gong2022}.  Algorithms have attempted to more explicitly account for long-range interactions by modulating convolutions with a mask defined by a local basis of Gaussians and a periodic basis of plane waves \cite{Cheng2021}, employing a unique global pooling scheme that could include additional context such as stoichiometry \cite{Louis2020}, or constructing additional features from the reciprocal representation of the crystal \cite{Yu2023}.  Other strategies have leveraged assumptions about the relationships among predicted variables, such as representing phonon spectra using a Gaussian mixture model \cite{Kong2022}. 

Given the promise and flexibility of graphical models, improving the data-efficiency, accuracy, generalizability, and scalability of these representations are active areas of research. While our previous discussion of structure-based material representations relied on the invariance of scalar properties to translation and rotation, this characteristic does not continue to hold for higher-order tensors. Consider a material with a net magnetic moment. If the material is rotated $180^{\circ}$ around an axis perpendicular to the magnetization, the net moment then points in the opposite direction. The moment was not invariant to the rotation but instead, transformed alongside the operation in an equivariant manner \cite{Batzner2021}. For a set of transformations described by group $G$, equivariant functions $f$ satisfy $ g*f(x)= f(g*x)$ for every input $x$ and every group element $g$ \cite{Batzner2021,Smidt2022}. Recent efforts have shown that by introducing higher-order tensors to node and edge features (\textbf{Figure \ref{graphical}d}) and restricting the update functions such that intermediate representations are equivariant to the group E3 (encompassing translations, rotations, and reflections in $R^{3}$), models can achieve state-of-the-art accuracy on benchmark datasets and even exhibit comparable performance to structural descriptors in low-data ($\sim$100 datapoints) regimes \cite{Thomas2018,Batzner2021,Painn2021}. Further accuracy improvements can be made by explicitly considering many-body interactions beyond edges \cite{Gasteiger2022a,Gasteiger2022,Batatia2022}. Such models, developed for molecular systems, have since been extended to solid-state materials and shown exceptional performance. Indeed, Chen et al. trained an equivariant model to predict phonon density of states and was able to screen for high heat capacity targets \cite{Chen2021a}, tasks identified to be particularly challenging for baseline CGCNN and MEGNet models \cite{Gong2022}. Therefore, equivariant representations may offer a more general alternative to the specialized architectures described above.

A major restriction of these graph-based approaches is the requirement for the positions of atomic species to be known. In general, ground-state crystal structures exhibit distortions that allow atoms to break symmetries, which are computationally modeled with expensive DFT calculations. Graphs generated from pristine structures lack representation of relaxed atomic coordinates (\textbf{Figure \ref{graphical}b}) and resulting model accuracy can degrade substantially \cite{Kolluru2022a, Gibson2022}. These graph-based models are therefore often most effective at predicting properties of systems for which significant computational resources have already been invested, thus breaking advice (3) from Section \ref{intro}. As a result, their practical usage often remains limited when searching broad regions of crystal space for an optimal material satisfying a particular design challenge. 

Strategies have therefore been developed to bypass the need for expensive quantum calculations and use unrelaxed crystal prototypes as inputs. Gibson et al. trained CGCNN models on datasets composed of both relaxed structures and a set of perturbed structures that map to the same property value as the fully relaxed structure. The data-augmentation incentivizes the CGCNN model to predict similar properties within some basin of the fully relaxed structure and was demonstrated to improve prediction accuracy on an unrelaxed test set \cite{Gibson2022}. Alternatively, graph-based energy models can be used to modify unrelaxed prototypes by searching through a fixed set of possibilities \cite{Schmidt2021} or using Bayesian optimization \cite{Zuo2021} to find structures with lower energy. Lastly, structures can be relaxed using cheap surrogate model (e.g. a force field) before a final prediction is made. The accuracy and efficiency of such a procedure will fundamentally rely on the validity and compositional generalizability of the surrogate relaxation approach \cite{Kolluru2022a}.

\section{CONSTRUCTING REPRESENTATIONS FROM STOICHIOMETRY}

The phase, crystal system, or atomic positions of materials are not always available when modeling materials systems, rendering structural and graphical representations impossible to construct. In the absence of this data, material representations can also be built purely from stoichiometry (the concentration of the constituent elements) and without knowledge of the geometry of the local atomistic environments. Despite their lack of structural information and apparent simplicity, these methods provide unique benefits for materials science researchers. First, descriptors used to form compositional representations such as common atomic properties (e.g. atomic radii, electronegativity) do not require computational overhead and can be readily found in existing databases \cite{Ward2016}. In addition, effective models can often be built using standard algorithms for feature selection and prediction that are implemented in freely available libraries \cite{Dunn2020}, increasing accessibility to non-experts when compared with structural models. Lastly, when used as tools for high-throughput screening, compositional models identify a set of promising elemental concentrations. Compared with the suggestion of particular atomistic geometries, stoichiometric approaches may be more robust, as they make weaker assumptions about the outcomes of attempted syntheses.

Compositional-based rules have long contributed to efficient materials design. Hume-Rothery and Linus Pauling designed rules for determining the formation of solid solutions and crystal structures that include predictions based on atomic radii and electronic valence states \cite{CallisterText,Pauling1929}. However, many exceptions to their predictions can be found  \cite{George2020}. 

Machine learning techniques offer the ability to discover and model relationships between properties and physical descriptors through statistical means. Meredig et al. demonstrated that a decision tree ensemble trained using a feature set of atomic masses, positions in the periodic table, atomic numbers, atomic radii, electronegativities, and valence electrons could outperform a conventional heuristic on predicting whether ternary compositions would have formation energies $<100$ meV/atom \cite{Meredig2014}. Ward et al. significantly expanded this set to 145 input properties, including features related to the distribution and compatibility of the oxidation states of  constituent atoms \cite{Ward2016}. Their released implementation, MagPie, can be a useful benchmark or staring point for the development of further research methods \cite{Dunn2020,Bartel2020,Stanev2018}. Furthermore, if a fixed structural prototype (e.g. elapsolite) is assumed, these stoichiometric models can be used to analyze compositionally-driven variation in properties \cite{Faber2016,Oliynyk2016}.

Even more subtle yet extremely expressive low-dimensional descriptors can be obtained by initializing a set with standard atomic properties and computing successive algebraic combinations of features, with each calculation being added to the set and used to compute higher order combinations in the next round. While the resulting set will grow exponentially, compressive sensing can then be used to identify the most promising descriptors from sets that can exceed $10^9$ possibilities \cite{Ouyang2018,Ghiringhelli2017}. Ghiringhelli et al. found descriptors that could accurately predict whether a binary compound would form in a zincblende or rocksalt structure \cite{Ghiringhelli2015}, and Bartel et al. identified an improved tolerance factor $\tau$ for the formation of perovskite systems \cite{Bartel2019} (\textbf {Table \ref{tab1}}). While these approaches do not derive their results from a known mechanism, they do provide enough interpretability to enable the extraction of physical insights for the screening and design of materials.

\begin{table}[h]
\label{tab1}
\tabcolsep7.5pt
\caption{Example Descriptors Determined through Compressive Sensing}
\label{tab1}
\begin{center}
\begin{tabular}{|c|c|c|}
\hline
Descriptor & Prediction & Variables\\
\hline

&&IP-Ionization Potential \\
$\frac{IP(B) - EA(B)}{r_p(A)^2}$ & Ordering in AB Compound &  EA-Electron Afinity\\
&&$r_p$-Radius of Maximum Density of p-Orbital \\
\hline

$ \frac{r_{X}}{r_{B}} - n_{A}(n_{A} - \frac{r_{A}/r_{B}}{ln[r_{A}/r_{B}]})$ & Stability of ABX$_{3}$ Perovskite & $n_{Y}$-Oxidation State of Y\\
&& $r_{y}$-Ionic Radius of Y\\

\hline
\end{tabular}
\end{center}
\end{table}

When large datasets are available, deep neural networks tend to outperform traditional approaches, and that is also the case for compositional representations. The size of modern materials science databases have enabled the development of information-rich embeddings that map elements or compositions to vectors as well as the testing and validation of deep learning models. Chemically meaningful embeddings can be constructed by counting all compositions in which that element appeared in the Materials Project \cite{Zhou2018} or learned through the application of natural language processing to previously reported results in the scientific literature \cite{Tshitoyan2019}. These data-hungry methods were able to demonstrate that their representations could be clustered based on atomic group \cite{Zhou2018} and could be used to suggest new promising compositions based on similarity with the best known materials \cite{Tshitoyan2019}. The advantages of training deep learning algorithms with large datasets are exemplified by ElemNet, which only uses a vector of fractional stoichiometry as input. Despite its apparent simplicity, when $>3,000$ training points where available, ElemNet performed better than a MagPie-based model at predicting formation enthalpies \cite{Jha2018}. 
 
 While the applicability of ElemNet is limited to problem domains with $O(10^3)$ datapoints, more recent methods have significantly reduced this threshold. ROOST \cite{Goodall2020} represented each composition as a fully-connected graph with nodes as elements, and properties were predicted using a message-passing scheme with an attention mechanism that relied on the stoichiometric fraction of each element. ROOST substantially improved on ElemNet, achieving better performance than MagPie in cases with only hundreds of training examples. Meanwhile, CrabNet \cite{Wang2021} forms element-derived matrices as a sum of embeddings of each element's identity and stoichiometric fraction. This approach achieves similar performance to ROOST by updating the representation using self-attention blocks. The fractional embedding can take log-scale data as input such that even dopants in small concentrations can have a significant effect on predictions. Despite the inherent challenges of predicting properties purely from composition, these recent and significant modeling improvements suggest that continued algorithmic development could be an attractive and impactful direction for future research projects.

Compositional models have the advantage that they can suggest new systems to experimentalists without requiring a specific atomic geometry and, likewise, can learn from experimental data without necessitating an exact crystal structure \cite{Zhang2021}. Owing to their ability to incorporate experimental findings into ML pipelines and provide suggestions with fewer experimental requirements (e.g. synthesis of a particular phase), compositional models have become attractive methods for materials design. Zhang el al. trained a compositional model using atomic descriptors on previous experimental data to predict Vicker's harness and validated their model by synthesizing and testing eight metal disilicides \cite{Zhang2021}. Oliynik et al. identified new Heusler compounds, while also verifying their approach on negative cases where they predicted a synthesis would fail \cite{Oliynyk2016}. Another application of their approach enabled the prediction of the crystal structure prototype of ternary compounds with greater than 96\% accuracy. By training their model to predict the probability associated with each structure, they were able to experimentally engineer a system (TiFeP) with multiple competing phases \cite{Oliynyk2017}.

While researchers have effectively implemented compositional models as methods for materials design, their limitations should be considered when selecting a representation for ML studies. Fundamentally, compositional models will only provide a single prediction for each stoichiometry regardless of the number of synthesizeable polymorphs. While training models to only predict properties of the lowest-energy structure is physically justifiable \cite{Tian2022}, extrapolation to technologically relevant meta-stable systems may still be limited. Additionally, graph-based structural models such as CGCNN \cite{Xie2018} or MEGNet \cite{Chen2019} generally outperform compositional models \cite{Bartel2020}. Therefore, composition models are most practically applicable when atomistic resolution of materials is unavailable, and thus structural representations cannot be effectively constructed. 

\section{DEFECTS, SURFACES, AND GRAIN BOUNDARIES}

Mapping the structure of small molecules and unit cells to materials properties has been a reasonable starting point for many applications of materials science modeling. However, materials design often requires understanding of larger length scales beyond the small unit cell, such as in defect and grain boundary engineering, and in surface science \cite{Artrith2019}. In catalysis, for example, surface activity is highly facet dependent and cannot be modeled using the bulk unit cell alone. It has been shown that the (100) facet of RuO$_2$, a state-of-the-art catalyst for the oxygen evolution reaction (OER), has an order of magnitude higher current for OER
than the active site on the thermodynamically stable (110) facet \cite{Rao2020}. Similarly, small unit cells are not sufficient for modeling transport properties, where size, orientation, and characteristics of grain boundaries play a large role. In order to apply machine learning to practical materials design, it is therefore imperative to construct representations that can characterize environments at the relevant length scales.

\begin{figure}[h]
\includegraphics[width=6in]{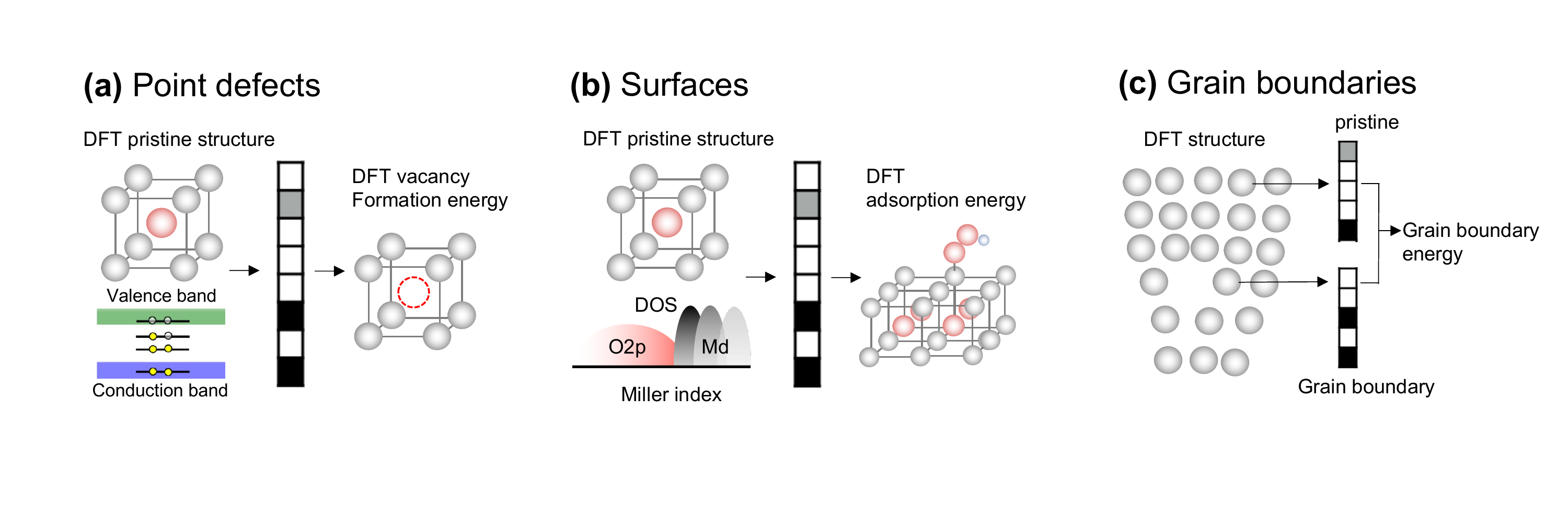}
\caption{\textbf{(a)} Point defect properties are learned from a representation of the pristine bulk structure and additional relevant information on conduction and valence band levels. \textbf{(b)} Surface properties are learned from a combination of pristine bulk structure representation, miller index, and density of states information. \textbf{(c)} Local environments of atoms near a grain boundary versus atoms in the pristine bulk are compared to learn grain boundary properties. \textbf{Figure \ref{defect}a} adapted from \cite{Varley2017}.}
\label{defect}
\end{figure}

Defect engineering offers a common and significant degree of freedom through which materials can be tuned. Data science can contribute to the design of these systems as fundamental mechanisms are often not completely understood even in long-standing cases such as carbon in steels \cite{Zhang2020}. Dragoni et al. \cite{Dragoni2018} developed a Gaussian Approximate Potential (GAP) \cite{Deringer2021} using SOAP descriptors for face-centered cubic iron that could probe vacancies, interstitials, and dislocations, but their model was confined to a single phase of one element and required DFT calculations incorporating $O(10^6)$ unique environments to build the interpolation. 

Considering that even a small number of possible defects significantly increases combinatorial complexity, a general approach for predicting properties of defects from pristine bulk structure representations could accelerate computation by orders of magnitude (\textbf{Figure \ref{defect}a}). For example, Varley et al. observed simple and effective linear relationships between vacancy formation energy and descriptors derived from the band structure of the bulk solid \cite{Varley2017}. While their model only considered one type of defect, their implementation limits computational expense by demonstrating that only DFT calculations on the pristine bulk were required \cite{Varley2017}. Structure- and composition-aware descriptors of the pristine bulk have additionally been shown to be predictive of vacancy formation in metal oxides \cite{Wan2021,Witman2022} and site/antisite defects in AB intermetallics \cite{Medasani2016}. To develop an approach that can be used over a broad range of chemistries and defect types, Frey et al. formed a representation by considering relative differences in characteristics (atomic radii, electronegativity, etc.) of the defect structure compared to the pristine parent\cite{Frey2020}. Furthermore, because reference bulk properties could be estimated using surrogate ML models, no DFT calculations were required for prediction of either formation energy or changes in electronic structure \cite{Frey2020}. We also note that in some cases it may be judicious to design a model that does not change significantly in the presence of defects. For these cases, representations based on simulated diffraction patterns are resilient to site-based vacancies or displacements \cite{Ziletti2018}.

Like in defect engineering, machine learning for practical design of catalyst materials requires representations beyond the single unit cell. Design of catalysts with high activity crucially depends on interactions of reaction intermediates with materials surfaces based on the Sabatier principle, which  argues that activity is greatest when intermediates are bound neither too weakly nor too strongly \cite{Medford2015}. From a computational perspective, determining absorption energies involves searches over possible adsorption active sites, surface facets, and surface rearrangements, leading to a combinatorial space that can be infeasible to exhaustively cover with DFT. Single dimension descriptors based on electronic structure have been established that can predict binding strengths and provide insight on tuning catalyst compositions, such as metal d-band center for metals \cite{Zhao2019} and oxygen 2p-band center for metal oxides \cite{Hwang2017}. Additional geometric approaches include describing the coordination of the active site (generalized coordination number in metals, adjusted generalized coordination number in metal oxides) \cite{Calle-Vallejo2015}. Based on the success of these simple descriptors, machine learning models have been developed to learn binding energy using the density of states and geometric descriptors of the pristine bulk structure as features (\textbf{Figure \ref{defect}b}) \cite{Fung2021}.

However, these structural and electronic descriptors are often not generalizable across chemistries \cite{Zhao2019,Back2019}, limiting the systems over which they can be applied and motivating the development of more sophisticated machine learning techniques. To reduce the burden on high-throughput DFT calculations, active learning with surrogate models using information from pure metals and active-site coordination has been used to identify alloy and absorbate pairs that have the highest likelihood of producing near-optimal binding energies \cite{Tran2018}. Furthermore, when sufficient data ($>10,000$ examples) is available, modifications of graph-convolutional models have also predicted binding energies with high accuracy even in datasets with up to 37 elements, enabling discovery without detailed mechanistic knowledge \cite{Back2019}. To generalize these results, the release of Open Catalyst 2020 and its related competitions \cite{Chanussot2021,Tran2022} has provided both over one million DFT energies for training new models and a benchmark through which new approaches can be evaluated \cite{Kolluru2022a}. While significant advancements have been made, state-of-the-art models still exhibit high-errors for particular absorbates and non-metallic surface elements, constraining chemistry over which effective screening can be conducted \cite{Kolluru2022a}. Furthermore, the complexity of the design space relevant for ML models grows considerably when accounting for interactions between absorbates and different surface facets \cite{Ghanekar2022}.

Beyond atomistic interactions, the mechanical and thermal behavior of materials can be significantly modulated by processing conditions and the resulting microstructure. Greater knowledge of local distortions introduced at varying grain boundary incident angles would give computational materials scientists a more complete understanding of how experimentally chosen chemistries and synthesis parameters will translate into device performance. Strategies to quantify characteristics of grain boundary geometry have included reducing computational requirements by identifying the most promising configurations with virtual screening \cite{Kiyohara2016}, estimating grain boundary free volume as a function of temperature and bulk composition \cite{Hu2020}, treating the microstructure as a graph of nodes connected across grain boundaries \cite{Dai2021,Reiser2022}, and predicting the energetics, and hence feasiblity, of solute segregation \cite{Huber2018}. While the previous approaches did not include features based on the constituent atoms and were only benchmarked on systems with up to three elements, recent work has demonstrated that the excess energy of the grain boundary relative to the bulk can be approximated across compositions with five variables defining its orientation and the bond lengths within the grain boundary (\textbf{Figure \ref{defect}c})\cite{Ye2022}. 

Further research has tried to map local grain boundary structure to function. Algorithmic approaches to grain boundary structure classification have been developed (see for example VoroTop \cite{Lazar2017}), but such approaches typically rely on expert users and do not provide a continuous representation that can smoothly interpolate between structures \cite{Priedeman2018}. To eliminate these challenges, Rosenbrock et al. proposed computing SOAP descriptors for all atoms in the grain boundary, clustering vectors into classes, and identifying grain boundaries through its local environment classes. The representation was not only predictive of grain boundary energy, temperature-dependent mobility, and shear coupling but also provides interpretable effects of particular structures within the grain boundary \cite{Rosenbrock2017}. A related approach computed SOAP vectors relative to the bulk structure when analyzing thermal conductivity \cite{Fujii2020}. Representations based on radial and angular structure functions can also quantify the mobility of atoms within a grain boundary \cite{Sharp2018}. When combined, advancing models for grain boundary stability as well as structure to property relationships opens the door for functional design of grain boundaries.

\section{TRANSFERABLE INFORMATION BETWEEN REPRESENTATIONS}

Applications of machine learning to materials science are limited by the scope of compositions and structures over which algorithms can maintain sufficient accuracy. Thus, building large-scale, diverse datasets is the most robust strategy to ensure trained models can capture the relevant phenomena. However, in most contexts, materials scientists are confronted with sparsely distributed examples. Ideally, models can be trained to be generalizable and exhibit strong performance across chemistries and configurations even with few to no data points in a given domain. In order to achieve this, representations and architectures must be chosen such that models can learn to extrapolate beyond the space observed in the training set. Effective choices often rely on inherent natural laws or chemical features that are shared between the training set and extrapolated domain such as physics constraints \cite{Batatia2022a,Axelrod2022}, the geometric \cite{Li2022, Harper2022} and electronic \cite{Unke2021, Husch2021} structure of local environments, and positions of elements in the periodic table \cite{Zheng2018, Feng2021}. For example, Li et al. were able to predict absorption energies on high entropy alloy surfaces after training on transition metal data by using the coordination number and electronic properties of neighbors at the active site \cite{Li2022}. While significant advancements have been made in the field, extrapolation of machine learning models across materials spaces typically requires specialized research methods and is not always feasible.

Likewise, it is not always practical for a materials scientist to improve model generality by just collecting more data. In computational settings, some properties can only be reliably estimated with more expensive, higher levels of theory, and for experimentalists, synthetic and characterization challenges can restrict throughput. The deep learning approaches that have demonstrated exceptional performance over a wide range of tests cases discussed in this review can require at least $~10^3$ training points, putting them seemingly out for the realm of possibility for many research projects. Instead, predictive modeling may fall back on identifying relationships between a set of human-engineered descriptors and target properties. 

Alternatively, the hidden, intermediate layers of deep neural networks can be conceptualized as a learned vector representation of the input data. While this representation is not directly interpretable, it must still contain physical and chemical information related to the prediction task, which downstream layers for the network utilize to generate model outputs. Transfer learning leverages these learned representations from task A and uses them in the modeling of task B. Critically, task A can be chosen to be one for which a large number of data points are accessible (e.g. prediction all DFT formation energies in the Materials Project), and task B can be of limited size (e.g. predicting experimental heats of formation of a narrow class of materials). In principle, if task A and task B share an underlying physical basis (the stability of the material), the features learned when modeling task A may be more informationally-rich than a human-designed representation \cite{Jha2019}. With this more effective starting point, subsequent models for task B can reach high accuracy with relatively few new examples.

The most straightfoward methods to implement transfer learning in the materials science community follow a common procedure: (1) train a neural network model to predict a related property (task A) for which $>O(1,000)$ data points are available (pretraining), (2) fix parameters of the network up a chosen depth $d$ (freezing), and (3) given the new dataset for task B, \textit{either} retrain the remaining layers, where parameters can be initialized randomly or from the task A model (finetuning), \textit{or} treat the output of the model at depth d as in input representation to another ML algorithm (feature extraction) \cite{Yamada2019, Gupta2021}. The robustness of this approach has been demonstrated across model classes including those using composition only (ElemNet \cite{Jha2019,Gupta2021}, ROOST \cite{Goodall2020}), crystal graphs (CGCNN) \cite{Lee2021}, and equivariant convolutions (GemNet) \cite{Kolluru2022}. Furthermore, applications of task B, range from experimental data \cite{Jha2019,Goodall2020} to DFT-calculated surface absorption energies \cite{Kolluru2022}. 

The sizes of the datasets for task A and task B will determine the effectiveness of a transfer learning approach in two ways. First, the quality and robustness of the representation learned for task A will increase as the number of observed examples (the size of dataset A) increases. Secondly, as the size of dataset B decreases, data becomes too sparse for a ML model to learn a reliable representation alone and prior information from the solution to task A can provide an increasingly useful method to interpolate between the few known points. Therefore, transfer learning typically exhibits the greatest boosts in performance when task A has orders of magnitude more data than task B \cite{Jha2019, Lee2021}.

In addition, the quality of information sharing through transfer learning depends on the physical relationship between task A and task B. Intuitively, the representation from task A provides a better guide for task B if the tasks are closely related. For example, Kolluru et al. demonstrated that transfer learning from models trained on the Open Catalyst Dataset \cite{Chanussot2021} exhibited significantly better performance when applied to absorption of new species than energies of less-related small molecules \cite{Kolluru2022}. While it is difficult to choose the optimal task A for a given task B a priori, shotgun transfer learning \cite{Yamada2019} has demonstrated that the best pairing can be chosen experimentally by empirically validating a large pool of possible candidates and selecting top performers.

 The depth $d$ from which features should be extracted from task A to form a representation can also be task dependent. Kolluru et al. provided evidence that to achieve optimal performance more layers of the network should be allowed to be retrained in step (3) as the connection between task A and task B becomes more distant \cite{Kolluru2022}. Gupta et al. arrived at a similar conclusion that the early layers of deep neural networks learned more general representations and performed better in cross-property transfer learning \cite{Gupta2021}. Inspired by this observation that representations at different neural network layers contain information with varying specificity to a particular prediction task, representations for transfer learning that combine activations from multiple depths have been proposed \cite{Kolluru2022, Chen2021}.

When tasks are sufficiently different, freezing neural network weights may not be the optimal strategy and instead representations for task B can include predictions for task A as descriptors. For instance, Cubuk et al. observed that structural information was critical to predict Li conductivity but was only available for a small set of compositions for which crystal structures were determined. By training a separate surrogate model to predict structural descriptors from composition and using those approximations in subsequent Li conductivity models, the feasible screening domain was expanded by orders of magnitude \cite{Cubuk2019}.  Similarly, Greenman et al. \cite{Greenman2022} used $O(10,000)$ TD-DFT calculations to  train a graph neural network whose estimates could be used as an additional descriptor for a model predicting experimental peaks in absorption spectra. Representations have also been sourced from the output of generative models. Kong et al. trained a Generative Adversial Network (GAN) to sample electronic density of states (DOS) given a particular material composition. Predictions of absorption spectra of a particular composition were improved by concatenating stoichiometric data with the average DOS sampled from the generative model \cite{Kong2021}.

\section{GENERATIVE MODELS FOR INVERSE DESIGN}

While, in principle, machine learning methods can significantly reduce the time required to compute materials properties, and material scientists can employ these models to screen for a set of target systems by rapidly estimating the stability and performance, the space of feasible materials precludes a naive global optimization strategy in most cases. Generative models including Variational Autoencoders (VAE) \cite{Kingma2013,GomezBombarelli2018}, Generative Adversarial Networks (GAN) \cite{Goodfellow2014,Nouira2018}, and diffusion models \cite{Shi2021,Ho2020} can be trained to sample from a target distribution and have proved to be capable strategies for optimization in high-dimensional molecular spaces \cite{GomezBombarelli2018,Schwalbe-Koda2020}. While some lessons can be drawn from the efforts of researchers in the computational chemistry community, generative models face unique challenges for proposing crystals \cite{Noh2020,Fuhr2022}. First, the diversity of atomic species increases substantially when compared with small organic molecules. In addition, given a composition, a properly defined crystal structure requires both the positions of the atoms within the unit cell as well as the lattice vectors and angles that determine the systems periodicity. This definition is not unique, and the same material can be described after rotations or translations of atomic coordinates as well as integer scaling of the original unit cell. Lastly, many state-of-the-art materials for catalysis (e.g. zeolites, metal organic frameworks) can have unit cells including $>100$ of atoms, increasing the dimensionality of the optimization problem \cite{Noh2020,Fuhr2022}. 

\begin{figure}[h]
\includegraphics[width=6in]{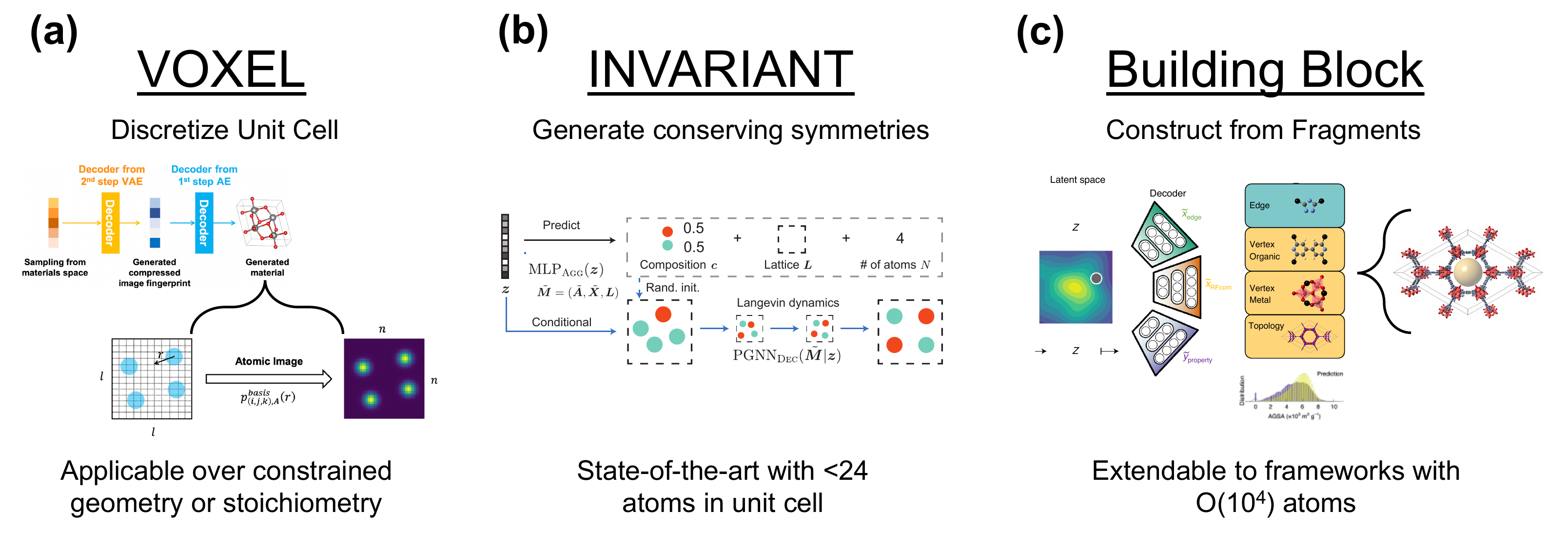}
\caption{Approaches for crystal structure generative models. \textbf{(Left)} Initial models based on voxel representations defined positions of atoms by discretizing space into finite volume elements but were not applied generally over the space of crystal structures \cite{Noh2019,Hoffmann2019,Court2020,Choubisa2020}. \textbf{(Center)} Restricting the generation process to be invariant to permutation, translation, and rotations, through an appropriately constrained periodic decoder (PGNN$_{Dec}$) results in sampling structures exhibiting more diversity and stability. \textbf{(Right)} When features of the material can be assumed, such as a finite number of possible topologies connecting substructures, the dimensionality of the problem can be substantially reduced and samples over larger unit cell materials can be generated. Figures on left, center, and right are adapted from \cite{Noh2019}, \cite{Xie2021}, and \cite{Yao2021}, respectively.}
\label{generative}
\end{figure}

One attempt to partially address the challenges of generative modeling for solid materials design is a voxel representation \cite{Noh2020}, in which unit cells are divided into volume elements and models are built using techniques from computer vision. Hoffman represented unit cells using a density field that could be further segmented into atomic species and was able to generate crystals with realistic atomic spacings. However, atoms could be mistakenly decoded into other species with nearby atomic number and most of the generated structures could not be stably optimized with a DFT calculation \cite{Hoffmann2019}. Alternate approaches could obtain more convincing results, but over a confined region of material space \cite{Ren2022}. iMatgen (\textbf{Figure \ref{generative}a}) invertibly mapped all unit cells into a cube with Gaussian-smeared atomic density and trained a VAE coupled with a surrogate energy prediction. The model was able to rediscover stable structures but was constrained over the space of Vanadium oxides \cite{Noh2019}. A similar approach constructed a separate voxel representation for each element and employed a GAN trained alongside an energy constraint to explore the phases of Bi-Se \cite{Long2021}. In order to resolve some of Hoffman el al's limitations, Court et al. \cite{Court2020} reduced segmentation errors by augmenting the representation with a matrix describing the occupation (0,1) of each voxel and a matrix recording the atomic number of occupied voxels.  Their model was able to propose new materials that exhibited chemical diversity and could be further optimized with DFT but restricted analysis to cubic systems. Likewise, compositions of halide perovskites with optimized band gaps could be proposed using a voxelized representation of a fixed perovskite prototype \cite{Choubisa2020}.

Voxel representations can be relaxed to continuous coordinates in order to develop methods that are more comprehensively applicable over crystal space. Kim et al. represented materials using a record of the unit cell as well as a point cloud of fractional coordinates of each element. The approach proposed lower energy structures than iMatgen for V-O binaries and was also applicable over more diverse chemical spaces (Mg-Mn-O ternaries) \cite{Kim2020a}. Another representation including atomic positions along with elemental properties could be leveraged for inverse design over spaces that vary in both composition and lattice structure. In a test case, the model successfully generated new materials with negative formation energy and promising thermoelectric power factor \cite{Ren2022}. While these models have demonstrated improvements in performance, they lack the translational, rotational, and scale invariances of real materials and are restricted to sampling particular materials classes \cite{Kim2020a, Xie2021}.

Recently, alternatives that account for these symmetries have been proposed. Fung et al. proposed a generative model for rotationally and translationally invariant atom-centered symmetry functions (ACSF) from which target structures could be reconstructed \cite{Fung2022}. Crystal Diffusion VAEs (\textbf{Figure \ref{generative}b}) leveraged periodic graphs and SE(3) equivariant message-passing layers to encode and decode their representation in a translationally and rotationally invariant way \cite{Xie2021}. They also proposed a two step generation process during which they first predicted the crystal lattice from a latent vector and subsequently sampled the composition and atomic positions through Langevin dynamics. Furthermore, they established well-defined benchmark tasks and demonstrated that for inverse design their method was more flexible than voxel models with respect to crystal system and more accurate than point cloud representations at identifying crystals with low formation energy.

Scaling solid-state generative modeling techniques to unit cells with $(10^4)$ atoms would enable inverse design of porous materials that are impossible to explore exhaustively but demonstrate exceptional technological relevance. Currently, due to the high number of degrees of freedom, sampling from these spaces requires imposing physical constraints in the modeling process. Such restrictions can be implemented as post-processing steps or integrated into the model representation. ZeoGAN \cite{Kim2020} generated positions of oxygen and silicon atoms in a 32x32x32 grid to propose new Zeolites. While some of the atomic positions proposed directly from their model violated conventional geometric rules, they could obtain feasible structures by filtering out divergent compositions and repairing bond connectivity through the insertion or deletion of atoms. Alternatively, Yao et al. designed geometric constraints directly into the generative model by representing Metal Organic Frameworks (MOFs) by their edges, metal/organic vertexes, and distinct topologies (RFcodes) (\textbf{Figure \ref{generative}c}) \cite{Yao2021}. Because this representation is invertible, all RFcodes correspond to a structurally possible MOFs. By training a VAE to encode and decode this RFcode representation, they demonstrated the ability to interpolate between structures and optimize properties. In general, future research should balance more stable structure generation against the possible discovery of new motifs and topologies.

\section{DISCUSSION}

In this review, we have introduced strategies for designing representations for machine learning in the context of challenges encountered by materials scientists. We discussed local and global structural features as well as representations learned from atomic-scale data in large repositories. We noted additional research that extends beyond idealized crystals to include the effects of defects, surfaces, and microstructure. Furthermore, we acknowledged that in practice the availability of data both in quality and quantity can be limited. We described methods to mitigate this including developing models based on compositional descriptors alone or leveraging information from representations built for related tasks through transfer learning. Finally, we analyzed how generative models have improved by incorporating symmetries and domain knowledge.
As data-based methods have become increasingly essential for materials design, optimal machine learning techniques will play a crucial role in the success of research programs. The previous sections demonstrate that the choice of representation will be among these pivotal factors and that novel approaches can open the door to new modes of discovery. Motivated by these observations, we conclude by summarizing open problems with the potential to have high impact on the field of materials design.

\subsection{Trade-offs of Local and Global Structural Descriptors}

Local structural descriptors including SOAP \cite{Bartok2013} have become reliable metrics to compare environments with a specific cutoff radius, and when properties can be defined through short-range interactions, have demonstrated strong predictive performance. Characterizing systems based off local environments allows models to extrapolate to cases where global representations may vary substantially (e.g. an extended supercell of a crystal structure)\cite{Musil2021} and enables highly-scalable methods of computation that can extend the practical limit of simulations to much larger systems \cite{Musaelian2022}. However, Unke et al. notes that the required complexity of the representation can grow quickly when modeling systems with many distinct elements and the quality of ML predictions will be sensitive to the selected hyperparameters, such as the characteristics distances and angles in atom-centered symmetry functions\cite{Unke2021b}. Furthermore, it is unclear if these high quality results extend to materials characteristics that depend strongly on long-range physics or periodicity of the crystal. On the other hand, recent global descriptors \cite{Sauceda2022}  can more explicitly model these phenomena, but have not exhibited the same generality across space groups and system sizes. Strategies exploring appropriate combinations of local and long-range features \cite{Grisafi2021} have the potential to break through these trade-offs to provide more universal models for material property prediction.

\subsection{Prediction from Unrelaxed Crystal Prototypes}

If relaxed structures are required to form representations, the space over which candidates can be screened is limited to those materials for which optimized geometries are known. Impressively, recent work \cite{Schaarschmidt2022,Lan2023} has shown that ML force-fields, even simple models with relatively high errors, can be used optimize structures and obtain converged results that are lower in energy than those obtained using VASP \cite{Kresse1993}. Their benchmarking on the OC20 \cite{Chanussot2021} dataset and lower accuracy requirements suggest that the approach could be generalizable across a wide-class of material systems and thus significantly expand the availability of structural descriptors. Similarly, Chen et al. demonstrated that a variant of MEGNET could perform high fidelity relaxations of unseen materials with diverse chemistries and that leveraging the resulting structures could improve downstream ML predictions of energy when compared with unrelaxed inputs \cite{Chen2022}. The strong performance of these approaches and their potential to significantly increase the scale and effectiveness of computational screening motivates high-value research questions concerning the scale of data sets required for training, the generalizabiltiy over material classes, and the applicability to prediction tasks beyond stability.

\subsection{Applicability of Compositional Descriptors}

Compositional descriptors are typically readily available as tabulated values, but even state-of-the-art models do not perform as well as the best structural approaches. However, there is some evidence that the scale of improvement when including structural information is property dependent. System energies can be conceptualized as a sum of site energies that are highly dependent on the local environment, and graph neural networks provide significantly more robust predictions of materials stability \cite{Bartel2020}. On the other hand, for properties dependent on global features such as phonons (vibrations) or electronic band structure (band gap) the relative improvement may not be as large \cite{Tian2022,Legrain2017,Zhuo2018}. Identifying common trends connecting tasks for which this difference is the least significant would provide more intuition on which scenarios compositional models are most appropriate. Furthermore, in some modeling situations, structural information is available but only over a small fraction of the dataset. To maximize the value of this data, more general strategies involving transfer learning \cite{Cubuk2019} or combining separate composition and structural models \cite{Stanev2018} should be developed.

\subsection{Extensions of Generative Models}

Additional symmetry considerations and the implementation of diffusion-based architectures led to generative models that improved significantly over previous voxel approaches. While this strategy is a promising direction for small unit cells, efforts pertaining to other parameters critical to material performance including microstructure \cite{Hsu2020}, dimensionality \cite{Song2020} and surfaces \cite{Liu2018} should also be pursued. In addition, research groups have side-stepped some of the challenges of materials generation by designing approaches that only sample material stoichiometry \cite{Dan2020}. While this strategy limits the full characterization of new materials through a purely computational pipeline, there may be cases where they are sufficient to propose promising regions for experimental analysis.

\section*{DISCLOSURE STATEMENT}
The authors are not aware of any affiliations, memberships, funding, or financial holdings that
might be perceived as affecting the objectivity of this review.

\section*{ACKNOWLEDGMENTS}
JD was involved in the writing of all sections. AT and MX collaborated on the writing and designed the figure for Atomistic Structure section, JK collaborated on the writing and designed the figure for the Periodic Graph section, JL collaborated on the writing and designed the figure for the Defects, Surfaces, and Grain Boundaries section. JP provided valuable insights for the organization and content of the article. RGB selected the topic and focus of the review, contributed to the central themes and context, and supervised the project. All authors participated in discussions and the reviewing of the final article. The authors would like to thank Anna Bloom for editorial contributions.

The authors acknowledge financial support from the Advanced Research Projects Agency–Energy (ARPA-E), US Department of Energy under award number DE-AR0001220. JD, MX and ART thank the National Defense Science and Engineering Graduate Fellowship, the National Science Scholarship from Agency for Science, Technology and Research, and Asahi Glass Company, respectively, for financial support. RGB thanks the Jeffrey Cheah Chair in Engineering.
\bibliographystyle{ar-style3.bst}
\bibliography{refs.bib}

\end{document}